\documentstyle[aps,prl]{revtex}
\begin{document}
\input psbox
\title{\bf  Anisotropic Sunyaev-Zel'dovich Effect }
\author{\bf Sujan Sengupta \\ 
Indian Institute of Astrophysics,
 Koramangala, Bangalore 560 034, India}
\date{\today}
\maketitle

\begin{abstract}
The spectral distortion in the Cosmic Microwave Background Radiation
(CMBR) caused from the inverse Compton scattering of low energy CMB photons by
thermally distributed hot electrons in massive clusters of galaxies -
better known as the thermal Sunyaev-Zel'dovich effect is usually
estimated by the analytical or semi-analytical solutions of
Kompaneets equation. These solutions are based on the assumption that the
scattering is isotropic, i.e., the intensity of the CMBR remains
angle-independent after scattering. In this letter, I present the solution
of the full radiative transfer equations by incorporating anisotropic
inverse Compton scattering of CMBR photons by relativistic electrons
and show that the thermal Sunyaev-Zel'dovich effect is strongly
angle-dependent. The spectral distortion in the CMBR alters significantly
in different angular directions. The amount of distortion increases
several times when the angle between the axis of symmetry and the
photon ray path is increased. It is pointed out that
for anisotropic scattering the Compton y parameter is angle dependent. Hence
it is important to consider the direction of the emergent radiation with
respect to the axis of symmetry in the electron rest frame while estimating
any cosmological quantity.

\end{abstract}


Inverse Compton scattering of the Cosmic Microwave Background Radiation (CMBR)
by hot intra-cluster gas - better known as the thermal Sunyaev-Zel'dovich
effect (TSZE) [1,2] results in a systematic transfer of photons from
the Rayleigh-Jeans to the Wien side of the spectrum causing a distortion
in the Planckian nature of the spectrum. The measurements of the effect yield
directly the properties of the hot intra-cluster gas, the total dynamical
mass of the cluster as well as the indirect information on the cosmological
evolution of the clusters. The effect is also used as an important tool to
determine the Hubble constant $H_{0}$ and the density parameter $\Omega_{0}$
of the Universe [3-9]. Recent interferometric imaging of the TSZE [10,11]
has been used to estimate the mass of various galaxy clusters which could
constrain the cosmological parameters of structure formation models.
Measurements of TSZE and the kinetic Sunyaev Zel'dovich effect (KSZE)
from the Sunyaev-Zel'dovich Infrared Experiment and OVRO/BIMA determine the
central Compton y parameter and constrain the radial peculiar velocity
of the clusters. In near future sensitive observations of the effect with
ground based and balloon-borne telescopes, equipped with bolometric multi-frequency
arrays, are expected to yield high quality measurements.
However, in modeling  the observational data, the analytical
solution of Kompaneets equation [12] with relativistic corrections are used
in general. The Kompaneets equation describes the isotropic intensity of the radiation
field after scattering. This means the expression used so far does not
describe the anisotropy or the angle-dependence of the radiation after
scattering. Hence the angle dependence of the spectral distortion in
CMBR is completely ignored in any theoretical discussion or interpretation
of the observed data in spite of the fact that scattering of an initially
isotropic radiation should become anisotropic. In this letter I show that
the thermal Sunyaev-Zel'dovich effect is strongly angle dependent.

   Chandrasekhar [13] has provided the radiative transfer equations for
the anisotropic Compton scattering by assuming that the electron energy
is much  less than the photon energy. On the other hand, for the
scattering of CMBR, the photon energy is much less than the electron energy.
In fact, in the TSZE, the electrons are considered to have relativistic
motion described by relativistic Maxwellian distribution. The relevant
radiative transfer equations that describe the anisotropic Compton
scattering of low energy photons can be written as [14-16]
\begin{eqnarray}
\mu\frac{\partial I(\mu,\nu,z)}{\partial \tau(z)}=-I(\mu,\nu,z) +
\frac{\omega_0}{2}\int^1_{-1}{\left[P_0-\frac{2kT_e}{m_ec^2}P_1+
\frac{2kT_e}{m_ec^2}\left(\nu^2\frac{\partial^2}{\partial\nu^2}-  
2\nu\frac{\partial }{\partial\nu}\right)P_2\right]I(\mu,\nu,z)d\mu'}.
\end{eqnarray}
Here $\omega_0$ is the albedo for single scattering,
$\mu=\cos\theta$ where $\theta$ is the angle between the axis of symmetry
(z axis) in the rest frame of electron gas and the ray path,
$\tau(z)$ is the total optical depth along the axis of symmetry
given by $d\tau(z)=\int{\sigma_Tn_e(z)dz}$,
$\sigma_T$ and $n_e$ being the Thomson scattering cross section and the electron
number density respectively.
 $k$, $c$, $m_e$, $\nu$ and
$T_e$ are Boltzmann constant, velocity of light, electron rest mass, frequency
of the photon  and the temperature of the electron respectively.
 $\omega_0=1$ for a purely
scattering medium. The phase functions $P_0$, $P_1$ and $P_2$ are given as :
\begin{eqnarray}
P_0(\mu,\mu')=\frac{3}{8}\left[3-\mu^2-\mu'^2(1-3\mu^2)\right],
\end{eqnarray}
\begin{eqnarray}
P_1(\mu,\mu')=\frac{3}{8}\left[1-3\mu'^2-3\mu^2(1-3\mu'^2)+2\mu^3\mu'(3-5\mu'^2)
+2\mu\mu'(3\mu'^2-1)\right],
\end{eqnarray}
and
\begin{eqnarray}
P_2(\mu,\mu')=\frac{3}{8}\left[3-\mu^2-\mu'^2+\mu\mu'(3\mu\mu'-5+3\mu^2+3\mu'^2-
5\mu^2\mu'^2)\right].
\end{eqnarray}

For an isotropic radiation field
 $\frac{1}{2}\int_{-1}^{1}{I(\mu',\nu)d\mu'}=I(\nu)$,
$\frac{1}{2}\int^1_{-1}{I(\mu',\nu)\mu'^2d\mu'}=\frac{1}{3}I(\nu)$ and
$\int^1_{-1}{I(\mu',\nu)\mu'd\mu'}=\int^1_{-1}{I(\mu,\nu)\mu'^3d\mu'}=0.$\footnote{
In Ref. [15] $I_0=\frac{1}{2}I$ should be $I=\frac{1}{2}I_0.$}
Therefore, for isotropic scattering with $s$ as the ray path that does not change
after scattering, equation (1) reduces to
\begin{eqnarray}
\frac{\partial I(\nu)}{\partial\tau(s)}=
\frac{2kT_e}{m_ec^2}\left(\nu^2\frac{\partial^2I(\nu)}{\partial\nu^2}-2\nu\frac{\partial I(\nu)
}{\partial\nu}\right)
\end{eqnarray}
which is well known as the Kompaneets equation [12] for low frequency.
Note that for plane parallel medium, the operator $\partial/\partial s$
becomes $\mu\partial /\partial z$ [13] which incorporates the change in
the ray path with respect to the z axis  after scattering.

The analytical solution of equation (5) provides the thermal component of
the distortion $\Delta I$ and is written as
\begin{eqnarray}
\Delta I(x)= I_0 y\frac{x^4e^x}{(e^x-1)^2}\left[\frac{x(e^x+1)}{e^x-1}-4\right]
(1+\delta_T)
\end{eqnarray}
where $x=\frac{h\nu}{kT_{CMBR}}$, $I_0=2(kT_{CMBR})^3/(hc)^2$.
 The term  $y=\frac{2kT_e}{m_ec^2}\int{\sigma_Tn_eds}$ 
is usually referred to as the Compton y parameter and $\delta_T$ is a relativistic
correction to the thermal effect [17] significant if $kT_e > 10 $KeV.

The above expression is usually used in order to estimate the TSZE. Clearly,
the angle dependence of the emergent intensity and hence the distortion is neglected
as the radiation field after scattering is assumed to remain isotropic.

Now equation (1) is a coupled integro partial differential equation and so
it cannot be solved analytically. I solve it numerically by discretization
method. In this method the medium is divided into several shells and the
integration is performed over two dimensional grids of angular and radial
points. For the angle integration I have adopted an eight point Gauss-Legendre
roots and weights. I have taken eighty frequency points with equal spacing.
For the initial condition, I have provided equal amount of intensity
corresponding to $T_{CMBR}=2.728$K at $\tau=0$ along all directions i.e.,
$I(\mu,\nu,\tau=0)=\frac{2h\nu^3}{c^2}(e^x-1)^{-1}$.  The code
is thoroughly tested for stability and flux conservation. The numerical results coincides
with the analytical solution given by equation (6) when the radiation is
made isotropic. On the other hand, if $P_1=P_2=0$, the results are well
matched with that presented in [13] for Thomson scattering.

Usually the isothermal $\beta$ models [18] is considered for the density
distribution of the clusters. The spherical  isothermal model density is
described by
\begin{eqnarray}
n_e(r)=n_0\left(1+\frac{r^2}{r_c^2}\right)^{-3\beta/2}
\end{eqnarray}
where the core radius $r_c$ and $\beta$ are shape parameters, $n_0$ is the
central electron density. However, for the present purpose it is sufficient
to consider an isothermal, homogeneous and plane parallel medium with a constant optical depth.
In the present work I have taken a constant value of $n_e=10^{-3}$ and
the size of the cluster is 4 Mpc. The electron temperature $T_e$ is taken
to be 1 KeV and 10 KeV. The results are presented graphically in Fig. 1
and in Fig. 2.

The spectral distortion for the isotropic case is characterized by three
distinct frequencies : the crossover frequency $x_0=3.83$ where the
TSZE vanishes; $x_{min}=2.26$ which gives the minimum decrement of the
CMB intensity and $x_{max}=6.51$ which gives the maximum distortion due
to this effect. The value of $x_0$ is however pushed to higher values
of $x$ with the increase in $T_e$ for the relativistic case.

  First of all there is no change in the values
of $x_0$, $x_{min}$ and $x_{max}$ in any direction even if the radiation
field is anisotropic. In order to show this,
the result for the isotropic scattering is also presented in Fig. 1 and in Fig. 2.
I have taken the ray path for the isotropic case along the axis of symmetry.
 The values of the three characteristic frequency points remain unchanged
because of the fact that the frequency
dependence of the distortion is independent of the angle dependence of
the distortion as can be seen from equation (1).

\begin{figure}[h]
\begin{center}
{\mbox{\psboxto(0cm;10cm){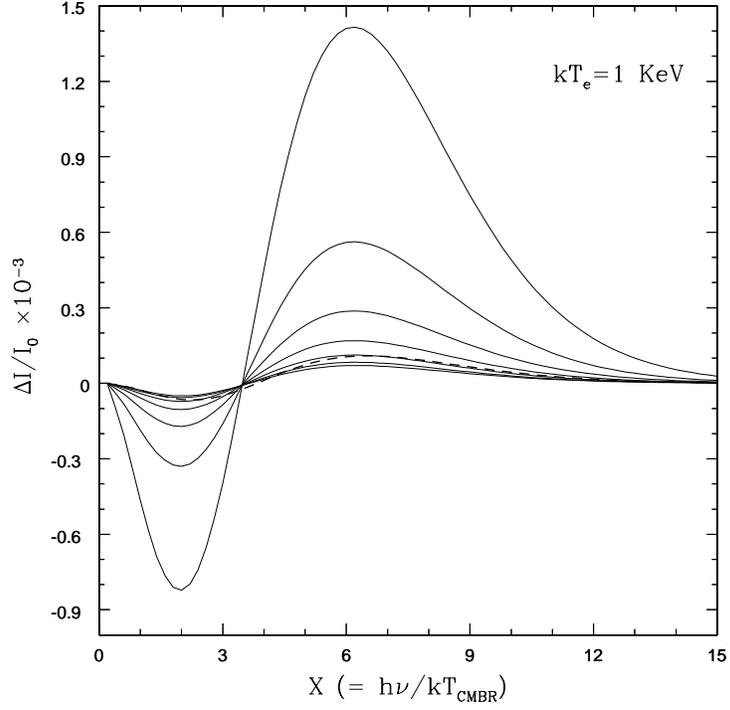}}}
\end{center}
\caption{
 Anisotropic Sunyaev-Zel'dovich effect with $kT_e=1$ KeV. From top
to bottom the solid lines represent the spectral distortion due to
anisotropic scattering  along the angular
directions (1)$\mu=0.1$, (2) $\mu=0.24$, (3) $\mu=0.4$, (4) $\mu=0.6$, 
(5)$\mu=0.76$, (6) $\mu=0.9$, (7) $\mu=0.98$. The dashed line
represent the distortion due to isotropic scattering with the ray path
being along the axis of symmetry.}
\end{figure} 

\begin{figure}[h]
\begin{center}
{\mbox{\psboxto(0cm;10cm){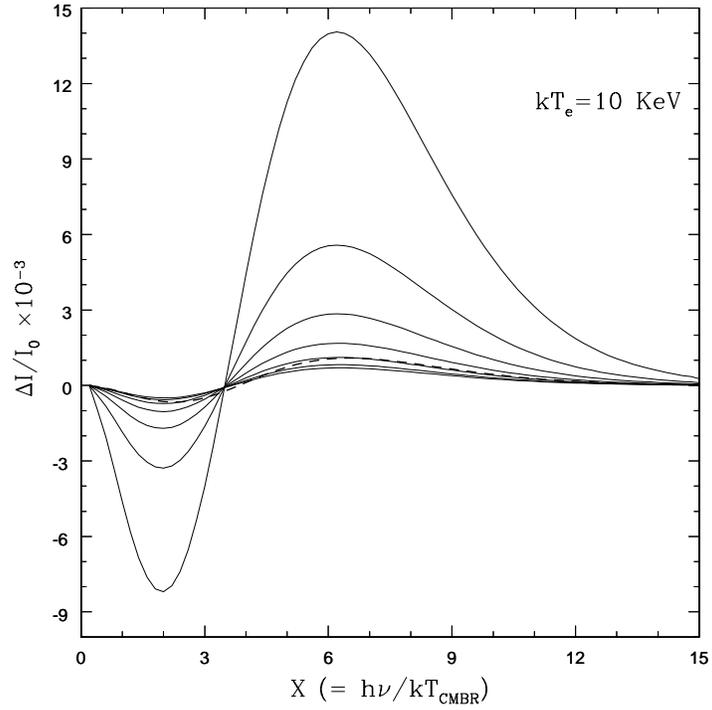}}}
\end{center}
\caption{
 Same as Fig.1 but with $kT_e=10$ KeV.}
\end{figure}

It is worth mentioning at this point that the photon changes its direction each
time it gets scattered. With respect to the z axis, the new ray
path becomes $s=z/\mu$. As a consequence, the optical depth through the
new ray path becomes $d\tau(z)/\mu$ where $d\tau(z)$ corresponds to the
optical depth along the z axis. Therefore, for
anisotropic scattering it is incorrect to take $\partial I/\partial \tau(s)$
instead of $\mu\partial I/\partial \tau(z)$ in the left hand side of
equation (1) because $\partial I/\partial \tau(s)$ describes the change
in the intensity along the same ray path $s$ before and after scattering.
Flux conservation confirms this fact. The Compton y parameter, for a single
anisotropic scattering in a plane-parallel medium, therefore, can be defined as
\begin{eqnarray}
y=\frac{1}{\mu}\frac{2kT_e}{m_ec^2}\int{\sigma_Tn_e(z)dz} 
\end{eqnarray}
where $\mu$ is now the angle between the incident direction to the emergent 
angle of the radiation.
In reality, however, the photon would suffer multiple scattering depending
on the density of the medium and hence would change its direction several times
before it emerges out.
If the medium is spherically symmetric then the situation becomes more
complicated. In plane parallel geometry the ray makes a constant angle $\theta$
with the normal while in the spherically symmetric geometry the angle made
by the ray direction and the radius vector changes constantly.

Now, the most interesting and important message conveyed by Fig.1 and
Fig. 2 is that the spectral distortion is strongly angle dependent. 
The degree of distortion in the CMBR due to TSZE changes drastically
with the change in the angular direction of the photons. 
This clearly demonstrates the fact that the TSZE produces anisotropic
distortion in the CMBR spectrum. 

  The degree of distortion increases as the angle between the emergent
intensity and the axis of symmetry increases. Therefore the minimum decrement and
maximum distortion of CMBR occur at $x=2.26$ and $x=6.51$ respectively for
the radiation emerging almost perpendicular to the axis of symmetry. For
a single scattering this can be explained easily. With the increase in $\theta$
and hence with the decrease in $\mu$ the Compton y parameter increases. As a
result the distortion increases when $\theta$ increases. For a plane 
parallel medium the Compton y
parameter becomes infinite when $\theta=\pi/2$ and hence the distortion
becomes infinite at $\theta=\pi/2$. It should be mentioned here that the results
presented in Fig. 1 and in Fig. 2 incorporate multiple scattering of photons
incident and emergent at any angular direction before and after scattering.
The radiation traversing at the opposite direction, i.e., the backscattered
radiation is also taken into care while solving equation (1) numerically.
The qualitative nature of the result
will not alter if we consider spherical symmetry except the fact that the
ray would peak with the radius vector towards the outer boundary of the
sphere. The results with spherical symmetry will be published in a
forthcoming paper.

Fig. 1 and Fig. 2 show that the spectral distortion due to isotropic
scattering matches with that due to anisotropic scattering when $\mu=0.76$.
Therfore the distortion due to anisotropic scattering would be less as
compared to that due to isotropic scattering if $\mu<0.76$. Now, the
spectral distortion along large angular direction should be irrelevant
from the observational point of view. But the important point to be noted
is that- if the effect is measured along the line of sight ( which can be
assumed as the axis of symmetry) then for the same value of the
Compton y parameter, the distortion due to anisotropic scattering would
be much less than that calculated under isotropic assumption. In other
words, the spectral distortion along the line of sight would result into
underestimation of the Compton y parameter if isotrpic scattering is
assumed.

Therefore it is extremely important to know the direction of the observed
intensity with respect to the axis of symmetry in the rest frame of the electron
gas while estimating the density of the clusters. The isotropic assumption
would either underestimate or overestimate it depending on the angular
direction of the emergent intensity of CMBR.

 In conclusion I would like to emphasize that the anisotropy in the CMBR,
induced by Compton scattering from hot electrons in the intra-cluster, not
only yields anisotropic distortion in the Planckian spectrum but also  would
results into polarization in the CMBR.  

I am thankful to P. Bhattacharjee for bringing my attention on this problem
and for discussions.

\end{document}